\documentclass[twocolumn,floatfix, showpacs]{revtex4}

\usepackage[final]{epsfig}
\usepackage{amsmath}
\begin{document}
 
\title{Charm energy loss and D-D correlations from a shower picture}
 
\author{Thorsten Renk}
\email{thorsten.i.renk@jyu.fi}
\affiliation{Department of Physics, P.O. Box 35, FI-40014 University of Jyv\"askyl\"a, Finland}
\affiliation{Helsinki Institute of Physics, P.O. Box 64, FI-00014 University of Helsinki, Finland}


\begin{abstract}
Measurements of mesons containing charm or bottom quarks at high transverse momentum ($P_T$) constitute an interesting set of probes for the nature of the interaction of hard partons with a QCD medium as created in ultrarelativistic heavy-ion (A-A) collisions. Not only can $D$ and $B$ mesons to a reasonable accuracy be assumed to represent the hadronization of a $b$ or $c$ quark after passage through the medium, i.e. they explicitly reflect quark-medium interaction, but also their interaction physics is expected to be different from light partons traversing a medium: The quark mass restricts radiation phase space, leading to a reduction of both vacuum and medium induced radiation, a phenomenon known as the 'dead cone effect'. Due to this difference in physics, heavy quark interaction with the medium is usually treated in the leading parton energy loss approximation, whereas the theory of light parton physics has moved to modeling the full medium-modified shower evolution. In this work, an attempt is made to create a more unified description of light parton and heavy quark physics at high $P_T$ by computing charm-medium interactions in terms of the nuclear suppression factor and back-to-back correlations using an in-medium shower evolution Monte Carlo (MC) code that is well tested for light parton physics.
\end{abstract}
 
\maketitle

\section{Introduction}

Measuring the energy loss of light partons (quarks and gluons) at high $p_T$ traversing a medium of thermalized Quantum-Chromodynamics (QCD) matter as created in ultrarelativistic heavy ion collisions has long been considered a valuable probe both of the macroscopic properties of the medium in terms of density distributions, and of the microscopical degrees of freedom \cite{radiative1,radiative2,radiative3,radiative4,radiative5,radiative6}. In the era of jet physics at the Large Hadron Collider (LHC), the emphasis has shifted from modeling leading parton energy loss towards a full in-medium evolution of the parton showers following a hard QCD process. Usually, MC codes are used to solve this problem \cite{JEWEL,YaJEM1,YaJEM2,Q-PYTHIA,MARTINI}.

In a parallel development, it was realized that the energy loss of high $p_T$ heavy quarks interacting with a medium would be different from that of light quarks as the quark mass restricts the kinematically available phase space for radiation, leading to a reduced medium-induced radiative energy loss, a phenomenon known as the 'dead cone effect' \cite{DeadCone}. The expectation was thus that the medium-modified spectrum of high $P_T$ electrons from A-A collisions, which chiefly results from the decay of $D$ and $B$ mesons, should show less suppression than the medium-modified spectrum of light hadrons, however experimentally about the same level of suppression was found \cite{HQexp1,HQexp2}. This has been termed the 'heavy quark puzzle' \cite{HQPuzzle}. As a solution for the puzzle, it was suggested that heavy quarks have an enhanced contribution of elastic reactions with the medium constituents when compared with light quarks, microscopically explained in terms of transient resonance formation of pre-hadrons even in a deconfined medium \cite{Resonant1,Resonant2,Resonant3} and phenomenologically implemented via a $K$-factor multiplying the elastic pQCD cross section \cite{Pheno1,Pheno2}.

A different contribution of elastic to radiative energy loss in the heavy quark sector as compared to the light parton case is an experimentally testable proposal. Previously it has been demonstrated that back-to-back hadron-hadron (h-h) correlations are an efficient tool to constrain the relative fraction of elastic energy loss in the light parton sector to be about 10\% \cite{ElCon1,ElCon2}. In a similar way, correlations of back-to-back $D$ mesons (D-D) could potentially be used to establish the relative contribution of elastic reactions to energy loss in the heavy quark sector as well.

This work aims at both treating light and heavy quarks on the same footing using a medium-modified shower evolution to compute the interaction of $c$-quarks with the medium and at predicting the order magnitude of the difference between h-h and D-D correlations using this common framework. In addition, the use of a full in-medium shower evolution allows to study also the question how the subleading hadrons in a heavy-quark induced shower are modified by the medium.

\section{The model}

A computation of a hard back-to-back correlation in a medium-modified shower framework requires three main building blocks: First, pQCD computation of the hard process itself, second a model for the evolution of the soft background medium and third link between the two, i.e. the medium-modified fragmentation pattern given the hard process and the soft medium. We present the results for h-h and D-D correlations side by side to illustrate where and how the physics of heavy quarks interacting with the medium is different.

In the following, we approximate the hard process itself by a leading order (LO) pQCD computation, as in the case of light parton production in principle supplemented by a $K$-factor to account effectively for higher order effects  and adjust the absolute normalization of hadron spectra to the data. Note that the relevant observables for medium modifications at high $P_T$, i.e. the nuclear modification factor

\begin{equation}
\label{E-RAA}
R_{AA}(P_T,y) = \frac{dN^h_{AA}/dP_Tdy }{T_{AA}({\bf b}) d\sigma^{pp}/dP_Tdy}.
\end{equation}

which is the ratio of the yield in A-A collisions as a function of transverse momentum $P_T$ and rapidity $y$, divided by the yield in p-p collisions scaled with the number of binary collisions, and the correlated yield suppression

\begin{equation}
\label{E-IAA}
I_{AA}(P_T,y) = Y_{med}(P_T,y)/Y_{vac}(P_T,y)
\end{equation}

which is the ratio of the in-medium per trigger conditional yield $Y_{med}(P_T,y)$ divided by the vacuum per-trigger yield $Y_{vac}(P_T,y)$, are both constructed in a way that a constant $K$-factor in the parton production cross section is canceled.

Higher order processes, both the splitting of a final state shower gluon $g \rightarrow c\overline{c}$ and charm production in initial state radiation which is subsequently scattered out of the nucleon wave function are known to contribute significanlty to the perturbative charm production, but are unlikely to produce a correlated hard back-to-back $c\overline{c}$ pair and are hence neglected in the following.

In LO pQCD, the production of two hard back-to-back partons $k,l$ 
is then described by
\begin{equation}
\label{E-2Parton}
  \frac{d\sigma^{AB\rightarrow kl +X}}{dp_T^2 dy_1 dy_2} \negthickspace 
  = \sum_{ij} x_1 f_{i/A}(x_1, Q^2) x_2 f_{j/B} (x_2,Q^2) 
    \frac{d\hat{\sigma}^{ij\rightarrow kl}}{d\hat{t}}
\end{equation}
where $A$ and $B$ stand for the colliding objects (protons or nuclei) and 
$y_{1(2)}$ is the rapidity of parton $k(l)$. The distribution function of 
a parton type $i$ in $A$ at a momentum fraction $x_1$ and a factorization 
scale $Q \sim p_T$ is $f_{i/A}(x_1, Q^2)$. The distribution functions are 
different for free protons \cite{CTEQ1,CTEQ2} and nucleons in nuclei 
\cite{NPDF,EKS98,EPS09}. The fractional momenta of the colliding partons $i$, 
$j$ are given by $ x_{1,2} = \frac{p_T}{\sqrt{s}} \left(\exp[\pm y_1] 
+ \exp[\pm y_2] \right)$.
 
The sum is taken over all relevant perturbative subchannels $\frac{d\hat{\sigma}^{ij\rightarrow kl}}{d\hat{t}}(\hat{s}, 
\hat{t},\hat{u})$ for the incoming partons $i$ and $j$ as a function of the parton Mandelstam variables $\hat{s}, \hat{t}$ and $\hat{u}$. In the case of $c\overline{c}$ productions, the dominant channel probed at LHC kinematics is $gg \rightarrow c\overline{c}$ with small additional contributions from $q\overline{q}\rightarrow c\overline{c}$

Eq.~(\ref{E-2Parton}) is in the following evaluated at midrapidity $y_1 = y_2 = 0$ and sampled using a MC code introduced in \cite{IAA_old} by first generating the momentum scale of the pair and then, in the case of light partons, the (momentum-dependent) identity of the partons. In the case of heavy quark production the parton identity is manifestly a $c\overline{c}$ pair (this implies that experimentally the presence of a heavy quark on both sides is tagged in some way). A randomly chosen $k_T$ with a Gaussian distribution of width 2.5 GeV is then added to the pair momentum. 

Under the assumption that the distribution of vertices follows binary collision scaling as appropriate for a LO pQCD calculation, the probability density to find a vertex in the transverse plane is

\begin{equation}
\label{E-Profile}
P(x_0,y_0) = \frac{T_{A}({\bf r_0 + b/2}) T_A(\bf r_0 - b/2)}{T_{AA}({\bf b})},
\end{equation}
where the thickness function is given in terms of Woods-Saxon distributions of the the nuclear density
$\rho_{A}({\bf r},z)$ as $T_{A}({\bf r})=\int dz \rho_{A}({\bf r},z)$ and $T_{AA}({\bf b})$ is the standard nuclear overlap function $T_{AA}({\bf b}) = \int d^2 {\bf s}\, T_A({\bf s}) T_A({\bf s}-{\bf b})$ for impact parameter ${\bf b}$. Each parton pair is placed at a probabilistically chosen vertex $(x_0,y_0)$ sampled from this distribution with a random orientation $\phi$ with respect to the reaction plane. Both partons are then propagated on eikonal paths through a hydrodynamical medium which has been shown to provide a constrained extrapolation from RHIC to LHC kinematics \cite{LHC_RAAold}. To take into account the fact that hard partons may still scatter from a medium not in thermal equilibrium (see discussion in \cite{HydSys}), interaction with the medium is tracked up to the hypersurface characterized by the temperature $T=130$ MeV.

The link between medium evolution and parton evolution is provided by the in-medium MC shower evolution code YaJEM in its version YaJEM-DE. This model is based on the PYSHOW code \cite{PYSHOW} which is part of PYTHIA \cite{PYTHIA} and is a well-tested tool to compute QCD showers in vacuum. YaJEM-DE simulates the evolution from a highly virtual initial parton to a shower of partons at lower virtuality in the presence of a medium down to a minimum scale $Q_0 = \sqrt{E/L}$ where $E$ is the energy of the shower initiator and $L$ is the in-medium pathlength \cite{YaJEM-D}.

YaJEM-DE is well tested against a number of different high $P_T$ observables both at RHIC and LHC kinematics, among them the dijet imbalance \cite{dijet1,dijet2} and the nuclear suppression factor for single hadrons and jets \cite{RAA-LHC} measured at LHC as well as dihadron \cite{ElCon2} and jet-hadron \cite{jet-h} correlations and the reaction plane angle dependence of the nuclear suppression factor $R_{AA}(\phi)$ \cite{YaJEM-D} as measured at RHIC.

The leading effects captured by YaJEM-DE are the modification of available radiation phase space by both a medium-induced correction to parton virtualities $\Delta Q^2$, leading to extra radiation, and a drag force term leading to an energy loss $\Delta E$ into non-perturbative modes of the medium, effectively reducing the radiation phase space. The relative contribution of these two mechanisms is determined by the data such that the drag force contributes $\sim 10\%$ to the total energy loss from the leading parton \cite{ElCon2}. A detailed description of the model can be found in \cite{YaJEM1,YaJEM2,ElCon2,YaJEM-D}. The dead cone effect is implemented inside PYSHOW (and YaJEM) as explicit parton-mass dependent limits on the phase space available for a branching.

Under the assumption that the relation between the medium thermodynamics in terms of the energy density $\epsilon$ and the transport coefficients $\hat{q}$ (determining the virtuality correction) and $\hat{e}$ determining the mean energy loss) can be written as 

\begin{equation}
\label{E-qhat}
\hat{q}[\hat{e}](\zeta) = K[0.1 K] \cdot 2 \cdot [\epsilon(\zeta)]^{3/4} (\cosh \rho(\zeta) - \sinh \rho(\zeta) \cos\psi)
\end{equation}

where $\psi$ is the angle between bulk medium flow and the parton direction, $\rho$ is the flow rapidity, and $K$ the adjustible parameter of the model regulating the overall strength of the parton-medium interaction, the medium induced perturbations to parton kinematics $\Delta Q^2$ and $\Delta E$ can be obtained by line integrals along the path $\zeta$ of each propagating intermediate virtual shower parton $a$ as

\begin{equation}
\label{E-Qgain}
\Delta Q_a^2 = \int_{\tau_a^0}^{\tau_a^0 + \tau_a} d\zeta \hat{q}(\zeta)
\end{equation}

and

\begin{equation}
\label{E-Drag}
\Delta E_a = \int_{\tau_a^0}^{\tau_a^0 + \tau_a} d\zeta \hat{e}(\zeta).
\end{equation}

Here, the time ordering of parton branching in the shower in terms of the virtual parton production time $\tau_a^0$ and its lifetime $\tau_a$ is obtained from the uncertainty relation and the virtual parton kinematics probabilistically as 

\begin{equation}
\label{E-Lifetime}
\langle \tau_a \rangle = \frac{E_a}{Q_a^2} - \frac{E_a}{Q_b^2} \quad \text{and} \quad P(\tau_a) = \exp\left[- \frac{\tau_a}{\langle \tau_a \rangle}  \right].
\end{equation}  

After the parton shower has been simulated to the lower scale $Q_0$, the Lund model \cite{Lund} is used to hadronize. The resulting event record is then investigated whether an observable-specific trigger condition is fulfilled, and if the condition is met, analyzed accordingly. The in-medium shower evolution of light parton and charm initiated showers is hence computed in an identical framework, the only difference being the constraints on the parton branching phase space as given by the quark mass. Note that these constraints are particularly effective for late branchings when the intermediate parton virtuality is small, for early times almost always $Q^2 \gg m_c^2$ is true, and hence quark masses do not lead to a sizeable correction.

\section{Effects of the charm mass on shower evolution}

Since QCD is flavour-blind, kinematic constraints set by the charm mass are indeed the only way the shower evolution of a charm quark can differ from that of a light quark. There are several places in which the quark mass is relevant: The kinematics of the shower evolution itself, the shower evolution time and the absence of thermally excited charm in the medium. Let us investigate these in turn.

In the PYSHOW framework, the QCD shower is treated as an interated series of $1\rightarrow 2$ splitting of a parent parton $a$ into two daughters $b,c$ where the energy is distributed as $E_a = z E_b + (1-z) E_c$ and the relevant virtuality scale (and hence the radiation phase space) decreases with every splitting as parametrized by $t = \log{Q^2}/{\lambda_{QCD}^2}$ until it reaches a set lower scale $Q_0$.
The differential splitting probability at a scale $t$ is then given by the splitting kernel $P_{a \rightarrow bc}(z)$ which can be computed in pQCD for the characteristic subprocess, integrated over the kinematically available range in $z$

\begin{equation}
I_{a\rightarrow bc}(t) = \int_{z_-(t)}^{z_+(t)} dz \frac{\alpha_s}{2\pi} P_{a\rightarrow bc}(z).
\end{equation}

where the kinematic limits available for $z$ depend on a combination of parent and daughter virtualities $Q_{abc}$ and masses $m_{abc}$ via $M_{abc} = \sqrt{m_{abc}^2 + Q_{abc}^2}$ as

\begin{widetext}
\begin{equation}
z_\pm = \frac{1}{2} \left( 1+ \frac{M_b^2 - M_c^2}{M_a^2}\pm \frac{|{\bf p}_a|}{E_a}\frac{\sqrt{(M_a^2-M_b^2-M_c^2)^2 -4M_b^2M_c^2}}{M_a^2} \right).
\end{equation}
\end{widetext}

It is this restriction of radiation phase space due to quark mass which makes the fragmentation in vacuum for heavy quarks much harder than for light quarks, and the same restriction is effective if the original hard virtuality $Q_a^2$ of a parton $a$ receives a medium-induced perturbation $\Delta Q_a^2$. 

The formation time of a shower partons can be estimated using the uncertainty principle as $\tau \sim E_a/M_a^2 = E_a / (m_a + Q_a)^2$. This relation implies that heavy quarks complete their virtuality evolution down to a non-perturbative scale parametrically faster than light partons. Since YaJEM is formulated as a perturbation of the virtuality evolution of shower partons, the code is conceptually not able to treat the interaction of on-shell quarks with the medium correctly. For 15 GeV $D$-meson production, the shower formation time can still be estimated to be above 2.2 fm, hence the shower still probes a sizable part of the densest evolution phase of the medium. Thus, with some potential caveats at lower energy, the formalism of YaJEM may well apply to charm quark evolution, but unless at very high energy it can be expected to fail for bottom quark showers as their formation time would be shorter than typically assumed formation times of the medium itself.

The third conceptual difference between heavy and light quark interactions with the medium is the absence of thermally excited $c\overline{c}$ pairs in the medium. While for a light quark propagating through a thermal QCD medium, parton-identity changing reactions like $q\overline{q} \rightarrow g g$ are possible with a scattering partner from the medium, no such partner exists in the heavy quark case and the channel hence does not contribute to heavy quark energy loss. However, as this channel turns out to be numerically very small in practice \cite{ElMC}, no strong effect for heavy quarks can be expected.

We may conclude from these considerations that YaJEM should be able to simulate charm quark showers propagating through a medium in a meaningful way, albeit with somewhat larger systematic uncertainties at lower $P_T$ than in the case of light parton showers.

\section{Results}

In the following, we use the setup described in \cite{RAA-LHC} where the parameter $K$ regulating the strength of the parton-medium interaction  is adjusted such as to give a good description of the nuclear suppression factor $R_{AA}$ for charged hadrons and jets at LHC kinematics. No additional free parameter is introduced for the computation of $c$-quark induced showers. 

\begin{figure}[htb]
\epsfig{file=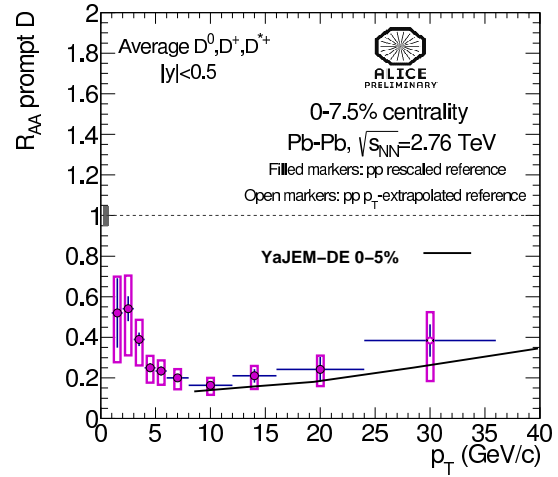, width=8cm}
\caption{\label{F-RAA} (Color online) Nuclear suppression factor of $D$-mesons as obtained by the ALICE collaboration \cite{ALICE-D} in 0-7.5\% central PbPb collisions at 2.76 ATeV, compared with YaJEM results for 0-5\% centrality. } 
\end{figure}

Fig.~\ref{F-RAA} shows the resulting nuclear suppression factor for $D$-mesons as compared with data obtained by the ALICE collaboration \cite{ALICE-D}. Within errors, the computation agrees well with the data and reproduces the slightly rising trend. There is a tendency of the computation to systematically undershoot the data, however note that the experimental centrality selection is somewhat larger than the one used for the calculation, hence a small upward shift of the model result is expected once this is taken into account.

This agreement with the data suggests that YaJEM indeed captures the essential physics of shower-medium interaction even for charm induced showers throughout the range of $\sim$8 to 35 GeV in transverse momentum. Armed with this result, we may now proceed to predict the strength of back-to-back correlation yields.

In the following, we compute the away side conditional yield suppression ratio $I_{AA}$ as a function of $P_T$ for identified $D$-mesons where $D$ generically stands for any of the hadrons $D^+, D^-, D^0, \overline{D^0}, D_s^+$ and $D_s^-$. As trigger condition, a $D$-meson in the range of 12 to 15 GeV is required. For the sake of illustration, the same computation is also carried out for charged hadrons and with RHIC and LHC kinematics side by side.

Based on the notions developed in \cite{ElCon1,ElCon2}, one can expect that the away side $I_{AA}$ is higher for incoherent scenarios of energy loss than for coherent ones. This can intuitively be understood using a simple picture of $L^2$ vs. $L$ weighting of the energy loss given a pathlength $L$. The nonlinear $L^2$ weighting leads to an increased surface bias for the vertices which lead to a trigger, i.e. partons being produced close to the surface propagating outward have a much higher chance to trigger as they lose little energy. However, this implies that away side partons have on average a long path through the medium which is weighted for energy loss purposes squared, thus such a scenario leads to a strong away side yield suppression. In an $L$-weighted scenario, the surface bias is less and hence the mean away side parton pathlength is also reduced, and in addition only a linear weighting factor is done, leading to a comparatively weak away side yield suppression. It can thus be expected that if the balance of elastic (incoherent) to radiative (coherent) energy loss for charm quarks is different, this would be reflected in the amount of the observed away side correlated yield suppression.

However, as discussed e.g. in \cite{Bayesian}, there is also a competing bias related to the medium-induced shift in the ratio of parton and trigger hadron energy. Thus, in order to gain an accurate understanding of the observable, the role of this kinematic bias needs to be investigated as well.

\begin{figure*}[htb]
\epsfig{file=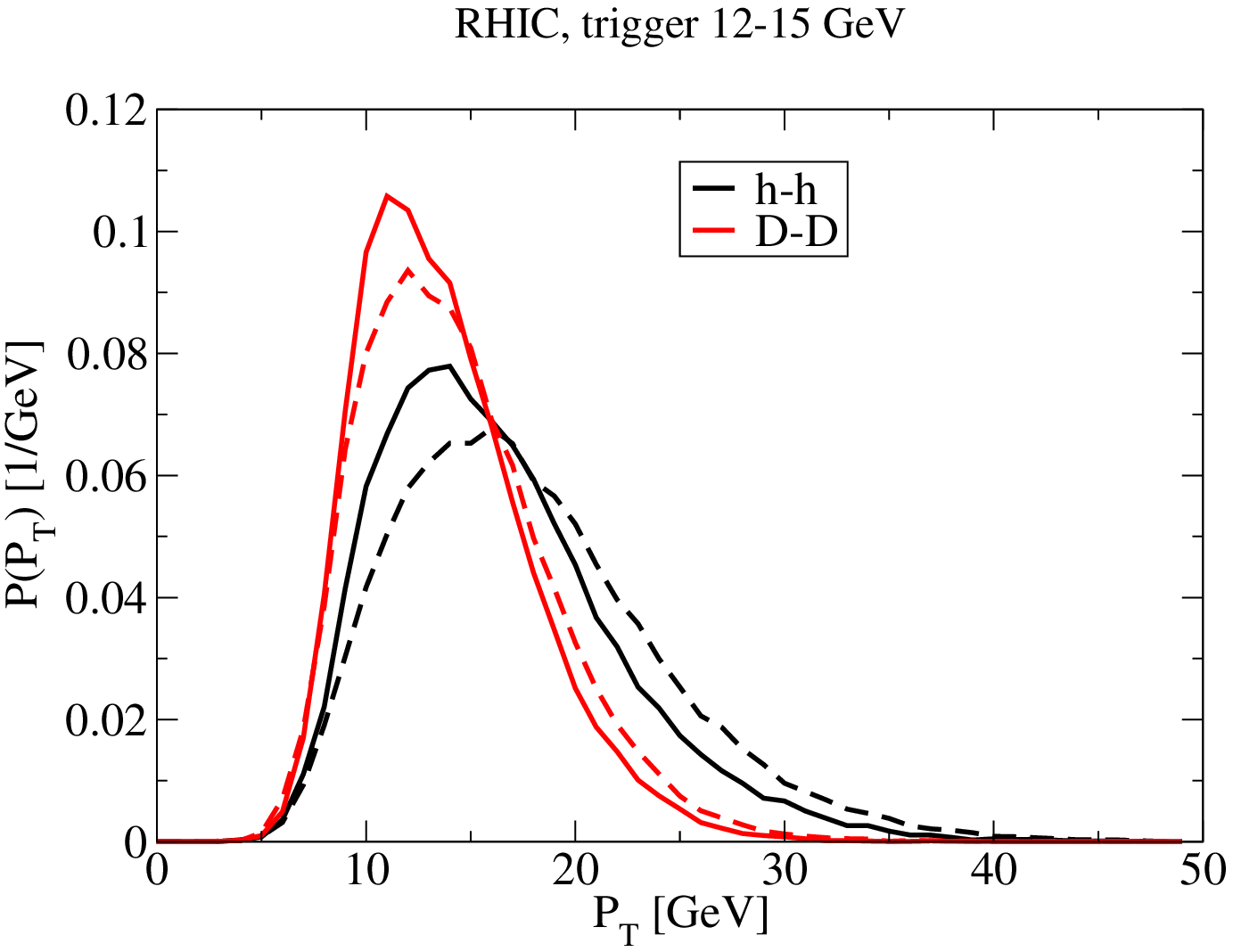, width=8cm}\epsfig{file=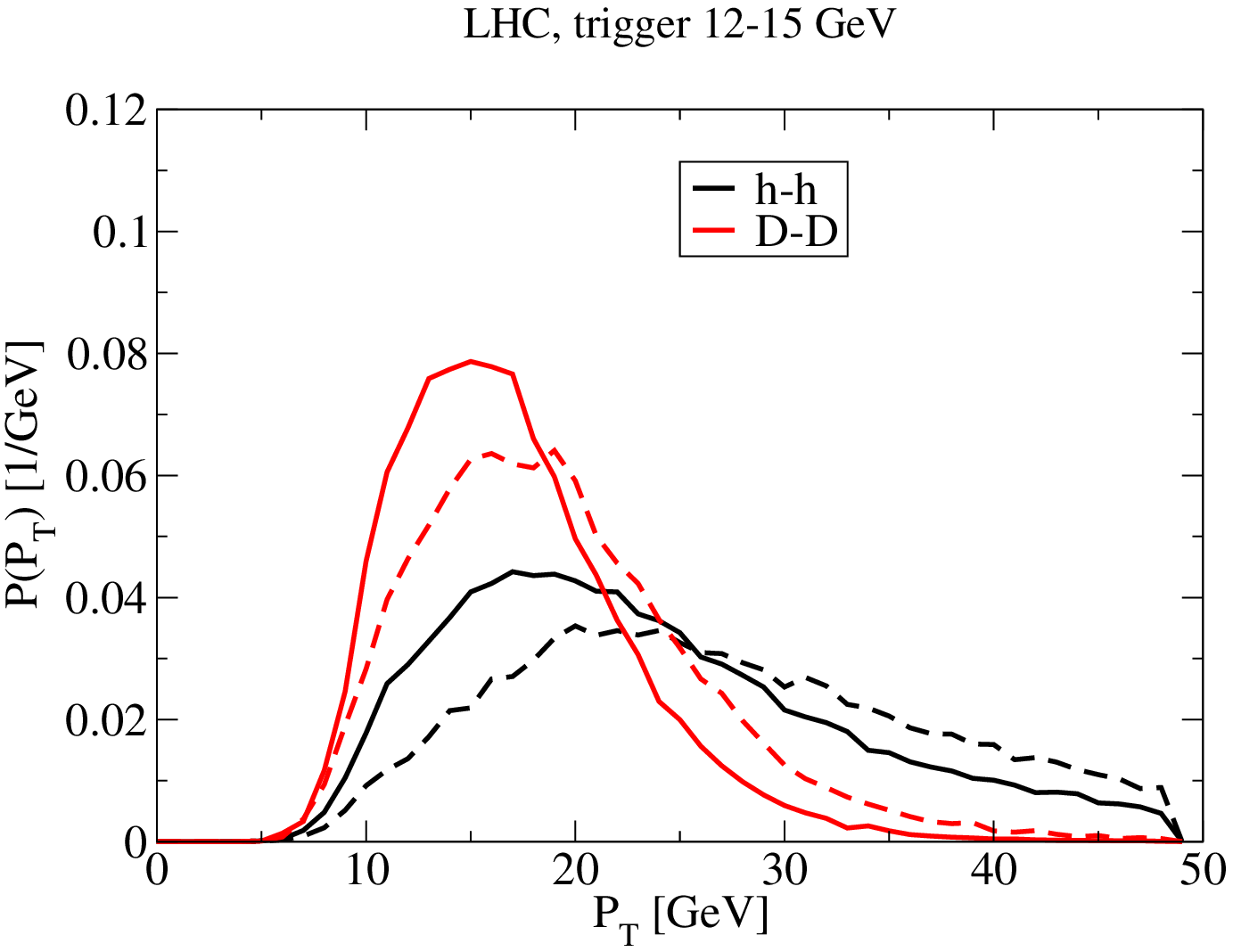, width=8cm}
\caption{\label{F-kinbias} (Color online) Distribution of parton momenta given a triggered hadron (or $D$-meson) in the range of 12-15 GeV for RHIC kinematics (left) and LHC kinematics (right). Shown are the results for vacuum (solid) and medium modified showers (dashed). } 
\end{figure*}

In Fig.~\ref{F-kinbias}, the relation between trigger kinematics and the kinematics of the underlying hard partonic event is shown in terms of the distribution of partonmomenta  given a trigger (either a $D$-meson or a charged hadron) in the range of 12-15 GeV, both for the vacuum and the medium modified case. Several general trends are clearly visible: First, the distribution of partons given a $D$-meson trigger is always narrower than the distribution of partons given a charged hadron trigger. This is another way of stating that the charm quark fragmentation function is harder, i.e. the $D$-meson is likely to carry  large fraction of the quark momentum. Second, the medium always shifts the distribution towards higher energies and also induced some broadening. This is a manifestation of energy loss on the trigger side --- since the trigger parton experiences some interaction with the medium, its energy before modification needs to be on average higher to produce a trigger hadron in the same range as a vacuum process. This is at the origin of the medium-induced kinematic bias \cite{Bayesian}. Third, at LHC kinematics all distributions are substantially wider than at RHIC kinematics. This can be understood in terms of the harder primary parton spectrum at LHC which leads to a reduced penalty for events in which an initially hard parton fragments with a low $z$ as compared to the RHIC case.

\begin{figure*}[htb]
\epsfig{file=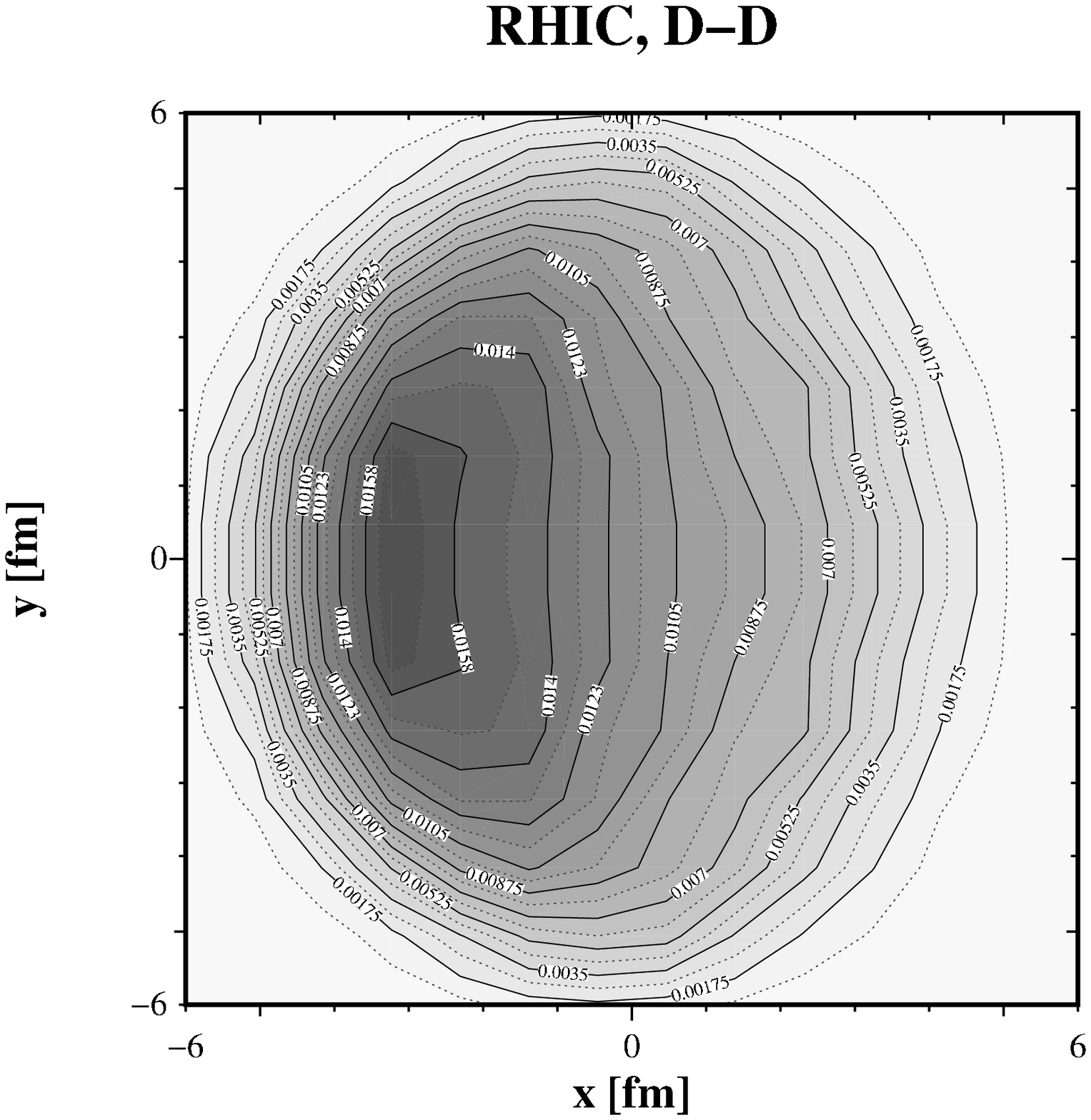, width=8cm}\epsfig{file=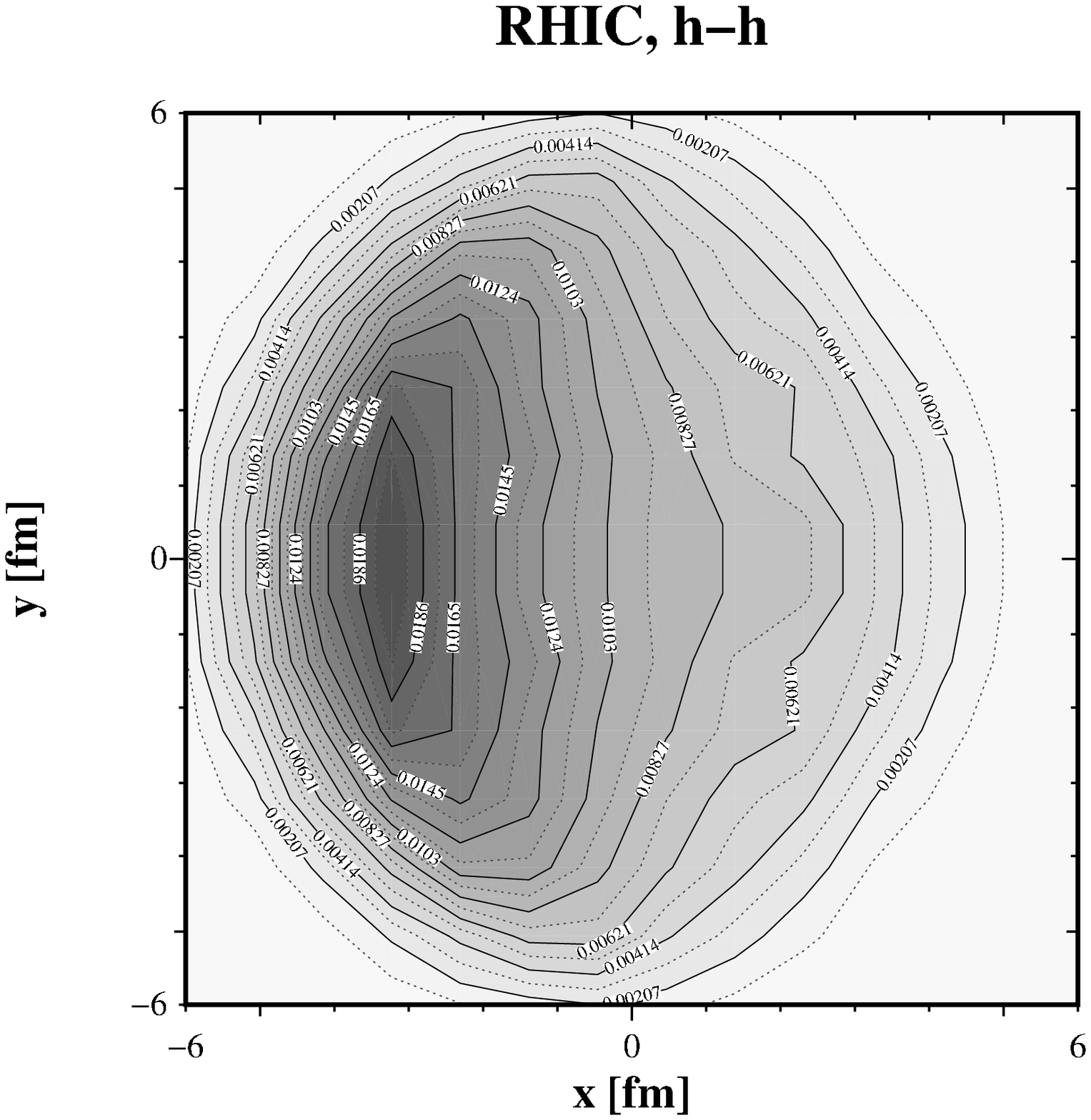, width=8cm}
\caption{\label{F-geoRHIC}Distribution of vertices in the transverse $(x,y)$ plane given a triggered $D$-meson (left) or charged hadron (right) for RHIC 200 AGeV 0-10\% central Au-Au collisions. }
\end{figure*}

\begin{figure*}[htb]
\epsfig{file=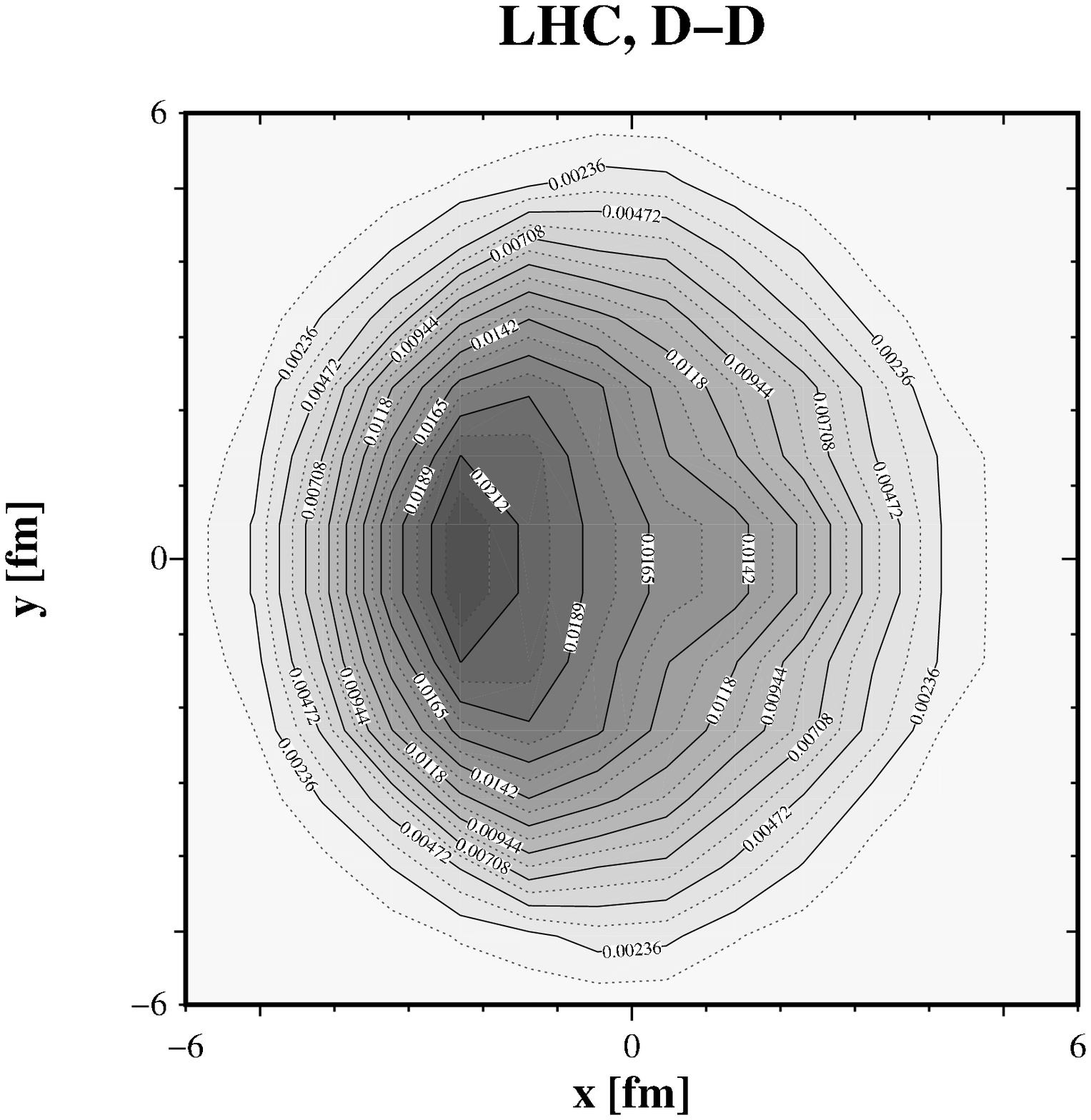, width=8cm}\epsfig{file=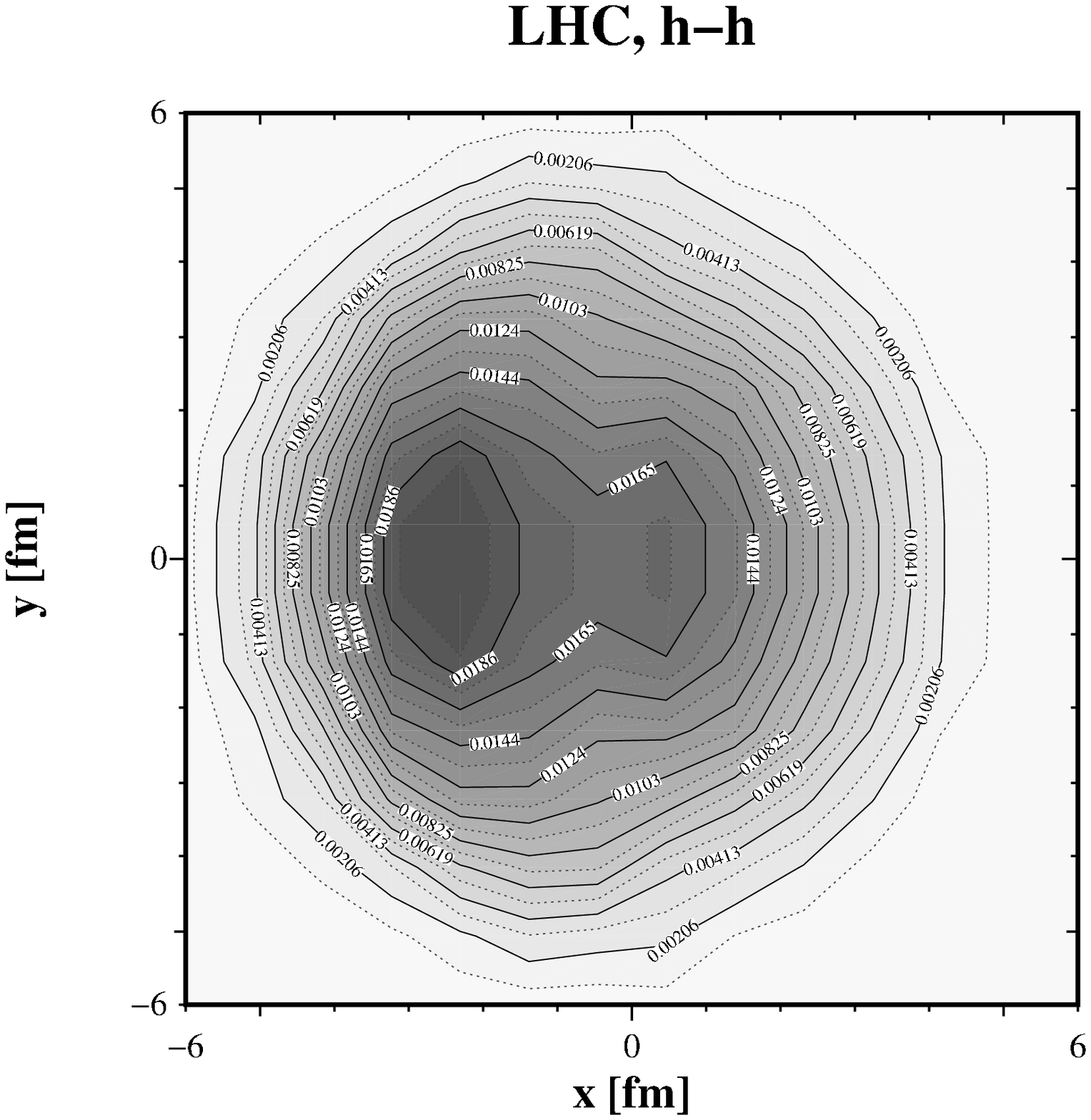, width=8cm}
\caption{\label{F-geoLHC}Distribution of vertices in the transverse $(x,y)$ plane given a triggered $D$-meson (left) or charged hadron (right) for LHC 2.76 ATeV central 0-10\% Pb-Pb collisions. }
\end{figure*}

The geometrical bias on the vertices of events leading to a trigger hadron can be seen from Fig.~\ref{F-geoRHIC} for RHIC kinematics and Fig.~\ref{F-geoLHC} for LHC kinematics. For RHIC conditions, the more pronounced surface bias for a charged hadron trigger as compared with a $D$ trigger, reflecting the reduced role of coherent radiation as a result of the dead cone effect is clearly seen. For LHC, no such trend is apparent from the figure. This can again be understood as a consequence of the harder primary parton spectrum which generically unbiases the geometry \cite{Bayesian}. Based on these findings, it is clear that one may not necessarily get to see the same trends of $I_{AA}$ at RHIC and LHC kinematics.

\begin{figure*}[htb]
\epsfig{file=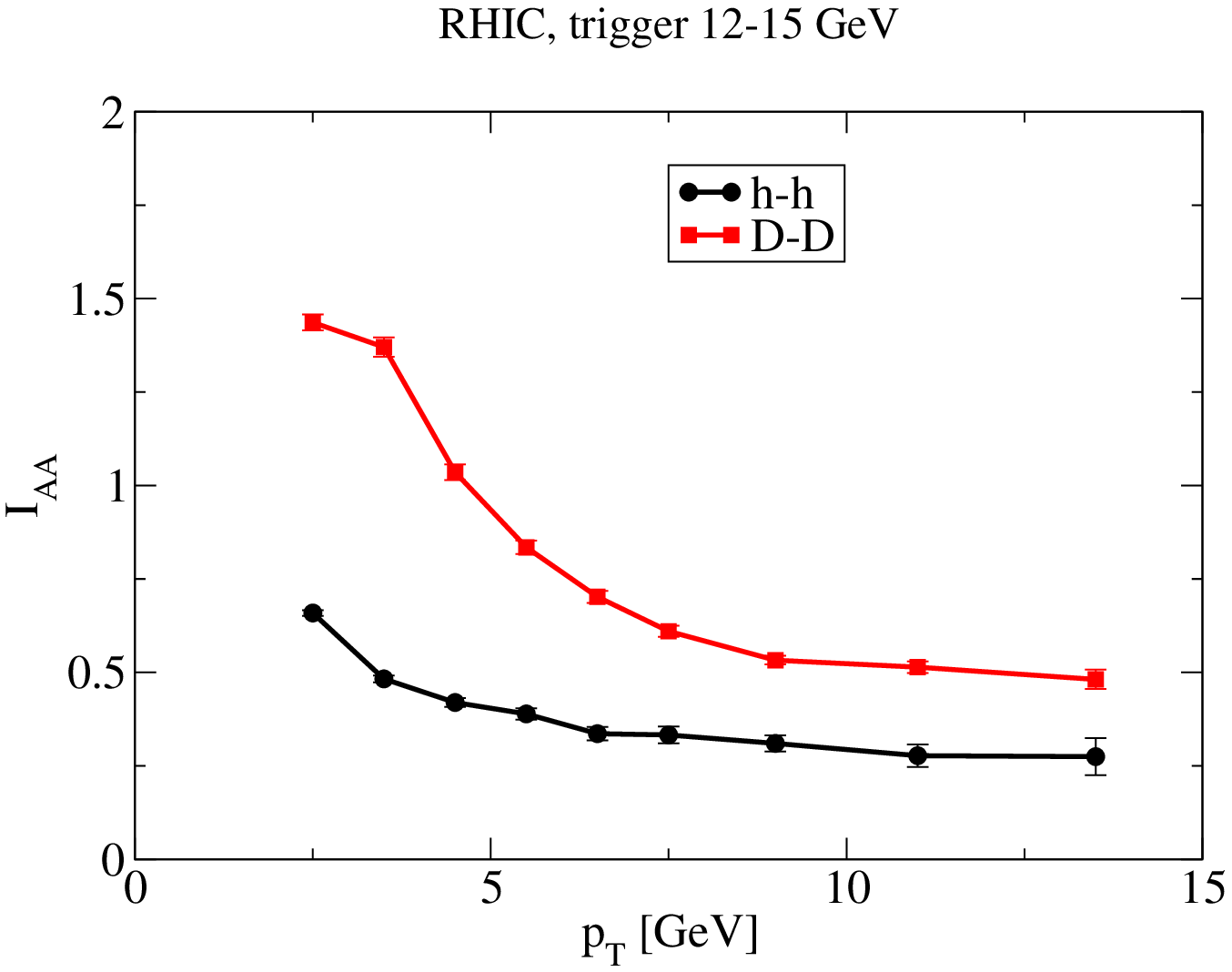, width=8cm}\epsfig{file=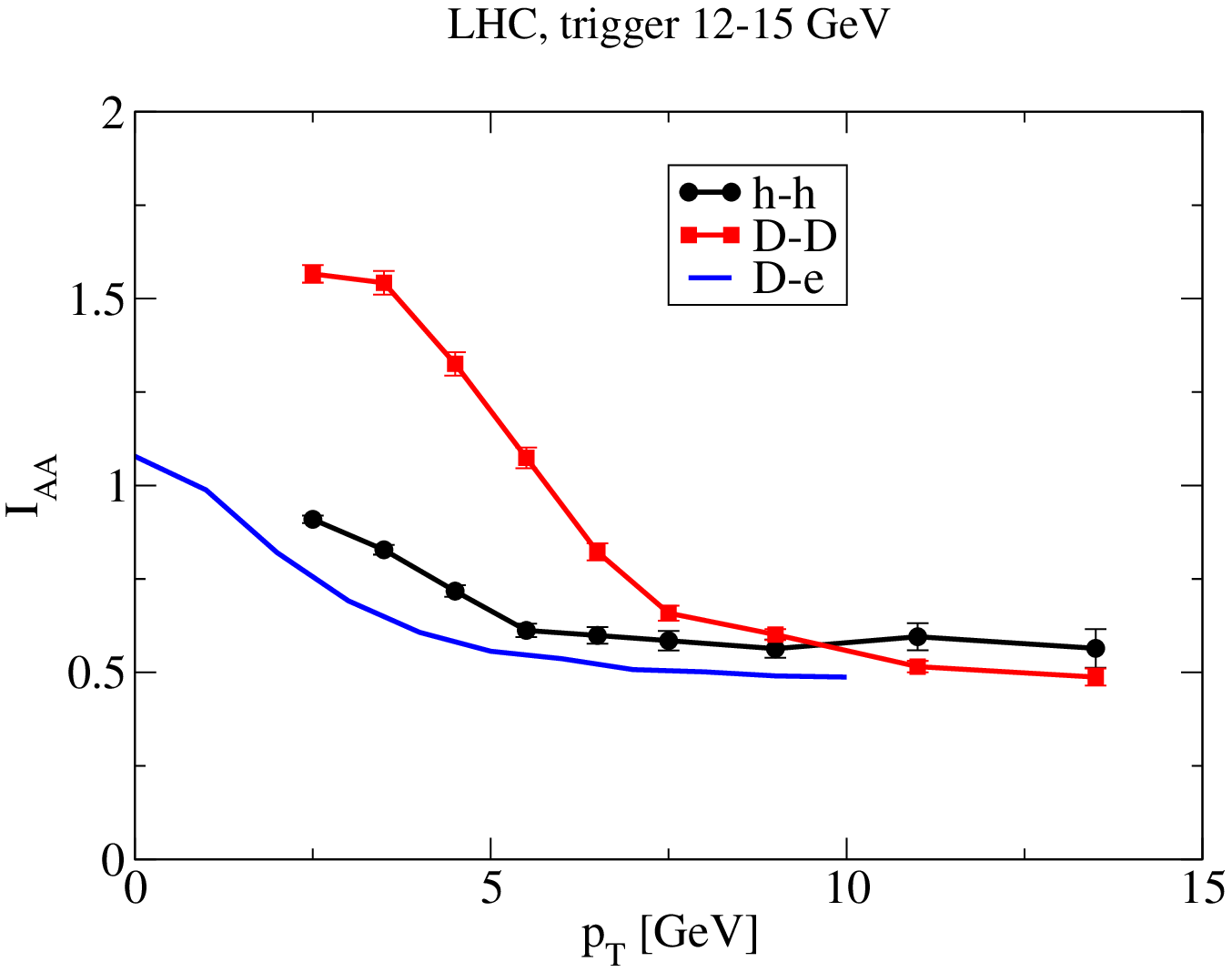, width=8cm}
\caption{\label{F-IAA}(Color online) Conditional away side yield suppression factor $I_{AA}$ for  RHIC 200 AGeV 0-10\% central Au-Au collisions (left) and  LHC 2.76 ATeV central 0-10\% Pb-Pb collisions (right)}
\end{figure*}

The resulting conditional yield ratios $I_{AA}$ for both D-D and h-h correlations at RHIC and LHC kinematics are shown in Fig.~\ref{F-IAA}. The RHIC result shows the expected ordering according to the importance of coherent radiative vs. incoherent elastic processes as a means to transport energy away from the leading parton. In agreement with the notion developed earlier, $I_{AA}$ for D-D correlations is substantially higher (almost a factor two) than for h-h correlations. 

The situation is significantly more difficult to interpret at LHC kinematics where in the high $P_T$ region both scenarios lead to an $I_{AA}$ of about 0.5, whereas there is a pronounced separation below about 7 GeV with the correlated yield in D-D again being significantly higher. In this case, the 'upturn point', i.e. the value of $P_T$ at which $I_{AA}$ crosses unity and changes from suppression to enhancement is the obvious signature distinguishing the two scenarios.

Let us first discuss the situation above 7 GeV where D-D and h-h results coincide. As stated earlier, the value of $I_{AA}$ is determined by a combination of geometry bias, i.e. the relative length of the typical near and away side pathlength, and by the kinematical shift induced by the medium. However, both biases do not act into the same direction: In a gedankenexperiment in which the medium density (and hence the medium induced energy loss) is increased, the geometry bias tends to decrease $I_{AA}$ whereas the kinematical bias tends to increase $I_{AA}$. The relative importance of these effects is set by the hardness of the primary parton momentum spectrum, as argued earlier a hard spectrum unbiases geometry, and thus the kinematic shift starts to dominate at increasing $\sqrt{s}$. At LHC kinematics, this happens by accident such as to cancel the effect of geometry in the same way between h-h and D-D correlations.

It can easily be seen even in an observable quantity that the agreement of h-h and D-D in the high $P_T$ region is an accident. If the same obseravble  $I_{AA}$ would be driven by the same physics, then a D-h correlation (where the h stands for any hadron \emph{except} the $D$-meson) could be expected to show also the same $I_{AA}$, as in this case the precise nature of the trigger would not matter.

\begin{figure}[htb]
\epsfig{file=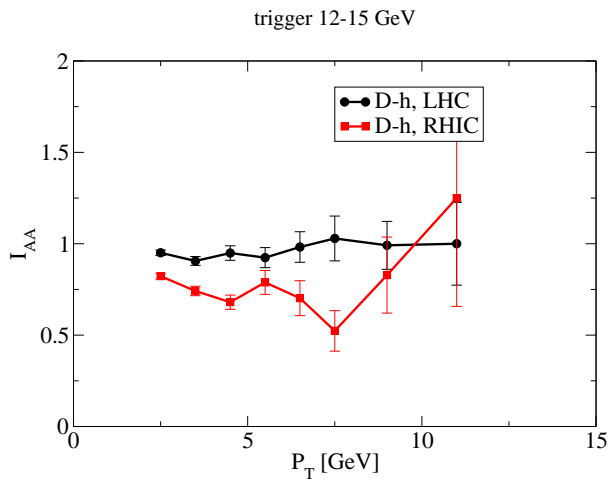, width=8cm}
\caption{\label{F-Dh} (Color online) Conditional away side yield suppression factor $I_{AA}$ for for RHIC and LHC kinematics for D-h correlations where the away side hadron is required explicitly not to be a $D$-meson. } 
\end{figure}

The result for D-h correlations is shown in Fig.~\ref{F-Dh}. Neither for RHIC nor for LHC kinematics is the $I_{AA}$ found to be even similar to the h-h or D-D result, clearly arguing that the precise nature of the trigger between a $D$-meson and a charged hadron causes a somewhat different bias structure. This clearly argues against the idea that measuring the same $I_{AA}$ for a different trigger object would be a sign of similar physics being probed.

Let us now focus on the region below 7 GeV where the most prominent difference between D-D and h-h is the location of the upturn point. In several light hadron observables, the upturn point has been seen to be remarkably independent of trigger $P_T$ scale at a location of around 3 GeV, such as in jet-hadron correlations at 10-15 and 20-40 GeV trigger energy \cite{jet-h-STAR} or the jet medium modified fragmentation function analysis for 120 GeV jets by CMS \cite{CMS-FF}. This important piece of data is reproduced by YaJEM in both cases \cite{jet-h,Bayesian}.

Within YaJEM, the upturn point is observed to be weakly trigger $P_T$ dependent and different for quark and gluon jets. The strong signal seen in the D-D case can hence largely be attributed to the fact that a back-to-back D meson shower is with good accuracy tagged to be a quark jet. This is not the case for h-h correlations where both at RHIC and LHC typically more than 70\% of the away side showers originate from gluons. 

While the fact that away side $I_{AA}$ in h-h and D-D correlations is in this computation expected to coincide at high $P_T$ for LHC kinematics, this is only unfortunate in terms of an easy and immediately evident interpretation of the results. Given the different \emph{a priori} distribution of partonic events in both cases, the fact that the conditional yield ratio is found to be the same nevertheless almost inevitably argues that the physics must be different, and such an argument can easily be supported by D-h correlations.

\section{Experimental feasibility}

Identifying a $D$-meson experimentally is difficult, and this may result in insufficient statistics to do D-D coincidences in practice. There are several possibilities how this problem might be lessened which all have specific advantages and disadvantages.

A D-H coincidence (where here H stands for all hadrons including the $D$-meson, distinct from D-h above where only hadrons different from the $D$ were correlated) would require an identified $D$-meson only on the near side.  To leading order pQCD, the leading away side hadron is statistically very likely to be a $D$-meson as the underlying partonic process leads to a recoiling $c\overline{c}$ pair and the charm fragmentation function into D is rather hard. However, unlike the D-D coincidence, the D-H coincidence suffers from NLO contributions from processes where charm appears only on the near side and the fragmenting away side may be a gluon. This could potentially erase the characteristic signatures of charm showers.

D-e correlations track the electron in the electromagnetic decay channel of the $D$-meson and hence ensure that a $c\overline{c}$ partonic structure has been present, however the kinematical information of the away side charm quark is much blurred by the decay kinematics. The combined fragmentation and decay function for the process $c\rightarrow e$ no longer shows a hard fragmentation in which the final object is kinematically well correlated with the shower initiating quark but qualitatively resembles more the soft fragmentation pattern of a gluon jet into hadrons. The result of this kinematical blurring is shown in Fig.~\ref{F-IAA}, right panel, for LHC kinematics where the away side $D$-meson spectrum has been decayed using PYTHIA \cite{PYTHIA}. The upturn point is shifted to lower $P_T$, and the final result resembles h-h correlations even more. This is not a surprise as the away side in h-h correlations is dominated by gluon jets, which, as stated above, have by accident a similar fragmentation pattern to electron production from fragmenting charm quarks. Given that the correlation structure is likely to be affected by background electrons at lower $P_T$, it can be seen from the figure that there is a clear benefit in having the complete $D$-meson kinematics on the away side.

Changing the trigger condition to an e-D, e-e or e-H correlation would break the kinematical relation between triggered object and charm quark momenta. As a consequence, there would be large fluctuations in parton energy given a triggered electron in a fixed momentum range. Such fluctuations are known to unbias the geometry (for reasons similar to why a harder parton spectrum leads to a weaker correlation between trigger and parton energy as discussed previously) and would hence tend to erase all the specific signals of heavy vs. light quark medium interactions in terms of coherence vs. incoherent elastic energy transfer. The specific bias structure of correlations sensitive to $c\overline{c}$ back to back events hence suggests that the best chance to probe the differences between light and heavy quark energy loss is to measure correlations where a $D$-meson could be identified at least on the trigger side.

\section{Conclusions}

As the results presented in this work show, correlation studies of $D$-meson triggered back-to-back correlations  have the potential to be sensitive to the specific differences in the light and heavy quark interactions with a QCD medium. The differences to h-h correlations driven by light parton physics predicted by this study are fairly robust and in some kinematical regions as large as a factor of two. The results however also indicate that understanding the detailed structure of the trigger bias in terms of kinematic shifts and geometry probed is mandatory and that a naive interpretation of the conditional yields is bound to fail.

The possibility to directly access the relative amount of elastic vs. radiative effects in the energy loss of the leading quark is exciting, as the strength of the elastic contribution provides direct insight into the nature of the degrees of freedom of the medium the hard parton scatters from. For instance, an ideal gas of massless perturbatively interacting partons can be shown to lead to a large elastic component incompatible with the data \cite{ElMCRP}, and dependent on assumptions about quasiparticle masses of the medium constituents, different hierarchies for the energy loss from light and heavy quarks can be expected \cite{MassHierarchy}.

In the past, high $P_T$ triggered back-to-back correlations have proven to be an extremely powerful toolkit to constrain the nature of the interaction of light partons with a QCD medium, hence applying similar techniques to high $P_T$ heavy quarks can therefore be expected to open a whole new avenue of experimentally probing the precise microscopical nature of the QCD medium produced in heavy-ion collisions.

\begin{acknowledgments}
 
This work is supported by the Academy researcher program of the
Academy of Finland, Project No. 130472. 
 
\end{acknowledgments}

\end{document}